\documentclass[12pt]{iopart}
\usepackage{iopams}
\usepackage{graphicx}
\expandafter\let\csname equation*\endcsname\relax
\expandafter\let\csname endequation*\endcsname\relax
\usepackage{amsmath}
\usepackage{amssymb}
\usepackage{cite}

\begin{document}

\title[First return times and the number of distinct sites visited by a random walk]
{The joint distribution of first return times and of the number of distinct sites 
visited by a 1D random walk before returning to the origin 
}

\author{Mordechai Gruda$^1$, Ofer Biham$^1$, Eytan Katzav$^1$ and Reimer K\"uhn$^2$}
\address{$^1$ Racah Institute of Physics, 
The Hebrew University, Jerusalem 9190401, Israel}
\address{$^2$ Department of Mathematics, King's College London, Strand, London WC2R 2LS, UK}
\eads{
\mailto{mordechai.gruda@mail.huji.ac.il},
\mailto{ofer.biham@mail.huji.ac.il}, 
\mailto{eytan.katzav@mail.huji.ac.il},
\mailto{reimer.kuehn@kcl.ac.uk}
}

\begin{abstract}

We present analytical results for the joint probability distribution 
$P(T_{\rm FR}=t,S=s)$
of first return (FR) times $t$ and of the number of distinct sites $s$  
visited by a random walk (RW) 
on a one dimensional lattice
before returning to the origin.
The RW on a one dimensional lattice is recurrent, namely the probability to return to the
origin is $P_{\rm R}=1$. However the mean 
$\langle T_{\rm FR} \rangle$
of the distribution $P(T_{\rm FR}=t)$ of first return times 
diverges. Similarly, the mean 
$\langle S \rangle$
of the distribution $P(S=s)$ of the number of distinct sites visited before
returning to the origin also diverges.
The joint distribution 
$P(T_{\rm FR}=t,S=s)$
provides a formulation that controls these divergences
and accounts for the interplay between 
the kinetic and geometric properties of first return trajectories.
We calculate the conditional distributions 
$P(T_{\rm FR}= t|S=s)$
and
$P (S=s|T_{\rm FR}=t)$.
Using moment generating functions and combinatorial methods, 
we find that the conditional expectation value 
of first return times of trajectories that visit $s$ distinct sites is
${\mathbb E}[T_{\rm FR} | S=s] = \frac{2}{3} (s^2+s+1)$,
and the variance is
${\rm Var}(T_{\rm FR} | S=s)=\frac{4}{45} (s-1)(s+2)(s^2+s-1)$.
We also find that in the asymptotic limit, the conditional expectation value of the
number of distinct sites visited by an RW that first returns to the origin at time $t=2n$ is
${\mathbb E}[S|T_{\rm FR}=2n] \simeq \sqrt{ \pi n }$,
and the variance is
${\rm Var}(S|T_{\rm FR}=2n) \simeq  \pi \left( \frac{\pi}{3} - 1 \right)  n$.
These results go beyond the important recent results of Klinger et al. 
[Klinger J, Barbier-Chebbah A, Voituriez R and B\'enichou O 2022 
{\it Phys. Rev. E} {\bf 105} 034116],
who derived a closed form expression for the generating function of the
joint distribution, but did not go further to extract an explicit expression for
the joint distribution itself.
The joint distribution provides useful insight on the efficiency of random 
search processes, in which the aim is to cover as many sites as possible in
a given number of steps.
A further challenge will be to extend this analysis to higher-dimensional lattices,
where the first return trajectories exhibit complex geometries.

\end{abstract}


\maketitle

\section{Introduction}

Random walks (RW) on discrete lattices
provide a useful tool for the analysis of
particle diffusion, diffusion-mediated reactions and search processes
\cite{Spitzer2001,Lawler2010}. 
Consider an RW on a hypercubic lattice in $d$ dimensions.
Starting at time $t=0$ from the origin $\vec r_0 = 0$,
at each time step $t \ge 1$ an RW hops
randomly to one of the $2 d$ neighbors of its present site.
The resulting trajectory takes the form
$\vec r_0 \rightarrow \vec r_1 \rightarrow \dots \rightarrow \vec r_t \rightarrow \dots$,
where $\vec r_{t}$ is the site visited at time $t$.
In some of the time steps the RW visits sites that
have not been visited before, while
in other time steps it revisits sites that have
already been visited at earlier times
\cite{Dvoretzky1951,Vineyard1963,Montroll1965}.
The latter possibility may take place either via a backtracking move
\cite{Tishby2017,Tishby2021},
in which the RW returns to the site occupied in the previous time step
or via a retroceding move
\cite{Tishby2021c}, in which, after a backtracking move, the RW
continues to hop backwards, revisiting sites it has visited three or more time
steps earlier.
In two and higher dimensional lattices, 
as well as in random and other complex networks,
an RW may also return to a previously visited site  
via a closed cycle, in a move referred to as retracing
\cite{Tishby2017,Tishby2021}.
The probability to visit a new, previously unvisited site is crucial for the survival of
foragers that consume the resources of the sites they visit.
The statistics of life expectancies of such foragers has recently been studied
on lattices of different dimensions
\cite{Benichou2014,Chupeau2016,Benichou2016}.

An RW starting from the origin at time $t=0$ may either return to the origin at a later
time or may wander farther and farther away  and never return to the origin.
An RW that returns to the origin with probability $P_{\rm R} = 1$ is called a recurrent RW,
while an RW that returns to the origin with probability $P_{\rm R} < 1$ is called a transient RW
\cite{Spitzer2001}.
In a seminal paper by G. P\'olya, published about 100 years ago,
it was shown that an RW on a $d$-dimensional hypercubic lattice is recurrent in 
dimensions $d=1,2$ and transient in dimensions $d \ge 3$
\cite{Polya1921}.
It was subsequently shown that RWs on Bethe lattices of degree $k \ge 3$
are always transient
\cite{Hughes1982,Cassi1989,Giacometti1995}.
This is in contrast to RWs on regular lattices and random networks of a finite size,
which are always recurrent
\cite{Kac1947,Tishby2021c}.
Moreover, in finite systems the distribution of first return times exhibits an
exponential tail
\cite{Harris1952}.
In finite systems an RW may eventually visit all the sites in the system.
The time at which the RW completes visiting all the sites in the system
at least once is called the cover time.
The distribution of cover times $P(T_{\rm C}=t)$ has been studied for finite
lattices and random networks of various geometries
\cite{Cooper2005,Tishby2022}.

For an RW starting from the origin $\vec r=0$ at time $t=0$, the
first return (FR) time $T_{\rm FR}$ 
is the first time at which the RW 
returns to the origin
\cite{Redner2001}.
The first return time varies between different instances of the RW
trajectory and its properties can be captured by a suitable distribution
\cite{Kostinski2016}.
The distribution of first return times  
is denoted by $P(T_{\rm FR}=t)$.
A more general problem involves the calculation of
the first passage (FP) time $T_{\rm FP}$,
which is the first time at which an RW starting from a given initial site $\vec r_0$ at time $t=0$
visits a specified target site  
\cite{Redner2001,Sood2005,Peng2021,Tishby2022b}
or a set of target sites
\cite{Redner2001,Baronchelli2006}.
The distribution of first passage times  from a given initial site to a given target site
is denoted by $P(T_{\rm FP}=t)$.

The distribution of first passage times is important in a variety of dynamical processes.
For example, consider an RW on a lattice, starting from a given initial site $\vec r_0$ at $t=0$ such that
there is a trap or sink at the origin, which locks and annihilates incoming RWs
\cite{Dayan1992}.
In this case, the fraction of RWs that survive the trap at time $t$ is given by the tail distribution of
first passage times, denoted by $P(T_{\rm FP} > t)$.
The distribution of first passage times on a one dimensional lattice can be used for the analysis
of the gambler's ruin problem, providing the probability that a gambler will run out of cash
after $t$ rounds
\cite{Coolidge1909,Feller1950}.

While the statistics of first passage times provides useful information about the kinetics
of random search processes, the number of distinct sites visited by an RW characterizes the geometry
of the domain it has covered within a given time window.
The mean number $\langle S \rangle_t$ of distinct sites visited by an RW 
on a one-dimensional lattice up to time $t$ satisfies 
$\langle S \rangle_t \propto \sqrt{t}$, while in two dimensions
it scales like $t/\ln t$ and in three and higher dimensions it scales linearly in $t$
\cite{Dvoretzky1951,Vineyard1963,Montroll1965}.
RWs on Bethe lattices satisfy
$\langle S \rangle \propto t$, namely they behave similarly to RWs on high dimensional lattices
\cite{Tishby2021c,Tishby2022,Debacco2015}.
Such behavior is maintained also for RWs on random networks of a finite size, as long as $t \ll N$,
where $N$ is the network size.

The interplay between the first passage time and the number of distinct sites
visited is captured by the joint probability distribution
$P(T_{\rm FP}=t,S=s)$ 
of first passage times $t$ and of the number of distinct sites $s$ visited by an RW
before reaching the target site for the first time.
In a remarkable recent paper, Klinger et al. studied the joint distribution 
$P(T_{\rm FP}=t,S=s)$
for a variety of random walk models
on a one-dimensional lattice
\cite{Klinger2022}.
Using the backward equation and a generating function formulation, they derived
the generating function of $P(T_{\rm FP}=t,S=s)$, from which the joint distribution 
can be obtained by differentiation.
However, they did not extract an explicit expression for the joint distribution itself.
Klinger et al. also showed that the interplay between the space and time variables 
can be captured by a scaling function.

In this paper we focus on the special case of first-return trajectories of RWs on a one dimensional lattice. 
We utilize recent developments regarding the combinatorial
properties of Dyck paths 
\cite{Flajolet2009,Hein2022,Merca2012,Merca2013,Merca2014,Fonseca2017}
to obtain analytical results for the joint distribution 
$P(T_{\rm FR}=t,S=s)$
of first return (FR) times $t$ and of the number of distinct sites $s$  
visited by an RW before returning to the origin.
Using the joint distribution 
we calculate the conditional distributions 
$P(T_{\rm FR}= t|S=s)$
and
$P (S=s|T_{\rm FR}=t)$.
Using moment generating functions, we obtain
closed-form expressions for
the conditional expectation value 
${\mathbb E}[T_{\rm FR} | S=s]$ 
of first return times of first return trajectories that visit $s$ distinct sites  
and the variance  
${\rm Var}(T_{\rm FR} | S=s)$.
We also obtain explicit expressions for the long-time limit of the conditional expectation value 
${\mathbb E}[S|T_{\rm FR}=2n]$
of the number of distinct sites visited by an RW that first returns to the origin at time $t=2n$  
and the variance  
${\rm Var}(S|T_{\rm FR}=2n)$.
The analytical results are
found to be in very good agreement with the results 
obtained from computer simulations.

The paper is organized as follows.
In Sec. 2 we present the random walk model on a one-dimensional lattice 
and review some useful results for
the distribution $P(T_{\rm FR} = t)$ of first return times.
In Sec. 3 we derive the joint distribution of first return times and the number of distinct sites
visited by an RW before returning to the origin.
In Sec. 4 we calculate the conditional distribution
$P(T_{\rm FR}=t | S=s)$ of first return times under the condition that the RW has visited
$s$ distinct sites before returning to the origin.
In Sec. 5 we calculate the mean and variance of
$P(T_{\rm FR}=t | S=s)$.
In Sec. 6 we calculate the conditional distribution
$P(S=s | T_{\rm FR}=t)$ of the number of distinct sites visited by an RW 
whose first return time to the origin is $t$.
In Sec. 7 we calculate the mean and variance of
$P(S=s | T_{\rm FR}=t)$.
The results are discussed in Sec. 8 and summarized in Sec. 9.
In Appendix A we consider different representations of the combinatorial factors
that account for the number of first return trajectories of length $2n$ in which the
RW visits $s$ distinct sites before returning to the origin.
In Appendix B we derive expressions for the moments of 
the conditional distribution $P(S=s|T_{\rm FR}=t)$
that are amenable to efficient numerical evaluation.

\section{First return times of a random walk on a one-dimensional lattice}

Consider an RW on a one dimensional lattice, starting from the origin at time $t=0$.
At each time step $t \ge 1$ the RW hops into a nearest neighbor site, either on the right or on the left,
with equal probabilities.
The location of the RW at time $t$ is denoted by $x_t$, which takes integer values.
At even time steps the RW visits even sites, namely $x_{2n} = \pm 2m$, 
while at odd time steps it visits odd sides, namely $x_{2n+1} = \pm (2m + 1)$, where $m=0, 1, 2, \dots, n$.
This implies that the RW may return to the origin only at even time steps.

The mean number of distinct sites visited by a random walk on a one-dimensional lattice
up to time $t$ is given by
\cite{Finch2003}

\begin{equation}
\langle S \rangle_t \simeq \sqrt{  \frac{8}{\pi} } \sqrt{t}.
\label{eq:Smean1D}
\end{equation}

\noindent
The  distribution of the number of distinct sites $s$ visited by an RW on a one-dimensional 
lattice up to time $t$ was recently studied
\cite{Regnier2022}.
The rate in which an RW discovers previously unvisited sites was studied in 
detail in higher dimensional lattices as well as in fractals and
disordered media and its universal features were revealed
\cite{Regnier2023}.
In the special case of random regular graphs, a closed-form expression 
was obtained for the distribution of the number of distinct sites $s$
visited by an RW up to time $t$
\cite{Tishby2022}.

Below we present the distribution of first return times.
In case that at time $t=1$ the RW hops to the right (left) hand side, the whole first
return trajectory is located on the positive (negative) half of the lattice.
Since the lattice is symmetric with respect to the origin and the RW hops
to the right or to the left with equal probabilities, there is a one to one 
correspondence between first return trajectories that reside on the positive and on the
negative sides.
For simplicity, in the analysis below, we assume that at $t=1$ the RW hops to
the right, namely $x_1=1$.

Since the first return time must be even, we use the parametrization
$P(T_{\rm FR}=2n)$, where $n$ is an integer.
The probability that the RW will 
first return to the origin at time $t=2n$,
is given by
\cite{Redner2001}

\begin{equation}
P(T_{\rm FR} = 2n) = 
C_{n-1} \left( \frac{1}{2} \right)^{2n-1},     
\label{eq:PTFRR}
\end{equation}

\noindent
where

\begin{equation}
C_{ k } = 
\frac{1}{k+1} \binom{2k}{k} 
\label{eq:Catalan}
\end{equation}

\noindent
is the Catalan number
\cite{Koshy2009,Stanley2015}.
The Catalan number $C_{k}$ counts
the number of discrete mountain ranges
of length $2k$
\cite{Deutsch1999,Audibert2010,Koshy2009}.
These mountain ranges, which are referred to as Dyck paths, 
describe RW trajectories on a one-dimensional lattice, starting from 
the origin at $t=0$ and returning to the origin at $t=2k$, where $x_{t} \ge 0$ for $0 \le t \le 2k$.
Thus, the Catalan number $C_k$ counts the number of RW trajectories on the non-negative part of
the lattice that return to the origin at time $t=2k$ (but not necessarily for the first time).
The number of such RW trajectories that return to the origin for the {\it first} time at $t=2k$
is given by $C_{k-1}$.
The probability that an RW will follow each one of these trajectories is $\left( \frac{1}{2} \right)^{2k-1}$.
In a first return trajectory the number of steps to the right 
direction is equal to the number of steps to the 
left direction.
Therefore, first return trajectories exist only for even values of the time $t=2n$.
In the trajectories considered here, the first step is always to the right,
while the last step is always to the left.
The other $t-2$ steps can be ordered in many different ways, as long as at each intermediate
time the number of steps to the right exceeds the number of of steps to the left by at least $1$.

The Catalan number can also be expressed in the form
\cite{Koshy2009,Stanley2015}

\begin{equation}
C_{ k } = 
\binom{2k}{k} - \binom{2k}{k+1}.
\label{eq:Catalan3}
\end{equation}

\noindent
The expression of Eq. (\ref{eq:Catalan3})  
clearly shows that the Catalan number must be an integer.
The first term on the right hand side of Eq. (\ref{eq:Catalan3})
is a central binomial coefficients that counts all the RW trajectories
that return to the origin after $2k$ steps. The second term on the right
hand side of Eq. (\ref{eq:Catalan3}) counts the RW trajectories that cross
the origin from the positive side to the negative side or vice versa before
returning to the origin at time $t=2k$.
Thus, the difference between these two terms accounts for the number of
RW trajectories that return to the origin after $2k$ steps without crossing
the origin up to that time (including trajectories that visit the origin at time
$t' = 2k' < 2k$ but do not cross to the other side).

In the long-time limit, applying the Stirling approximation on Eq. (\ref{eq:PTFRR})  
yields 
$P(T_{\rm FR} = 2n) \propto  n^{-3/2}$.
This is consistent with earlier results showing that the mean first return time 
$\langle T_{\rm FR} \rangle$ 
diverges
\cite{Polya1921}.
This observation motivates the consideration of methods to control the divergence
of the mean first return time.
A useful formulation that provides such control involves the joint distribution of 
the first return time and of the number of distinct sites visited by the RW before 
returning to the origin.
This formulation is developed in the next section below.

\section{The joint distribution of first return times and of the number of distinct sites visited by an RW
before returning to the origin}

First return trajectories of RWs on a one dimensional lattice exhibit an interplay between
space and time. For example, RWs that reach large distances from the origin
(and thus visit a large number of distinct sites) are also likely
to take a long time to return to the origin, and vice versa.
The detailed correlation between the first return time $t$ and the number of distinct sites
$s$ visited by an RW before returning to the origin for the first time is captured by the
joint probability distribution
$P(T_{\rm FR}=t,S=s)$.
Below we derive an exact analytical expression for this joint distribution and explore its properties.

\begin{figure}
\centerline{
\includegraphics[width=10cm]{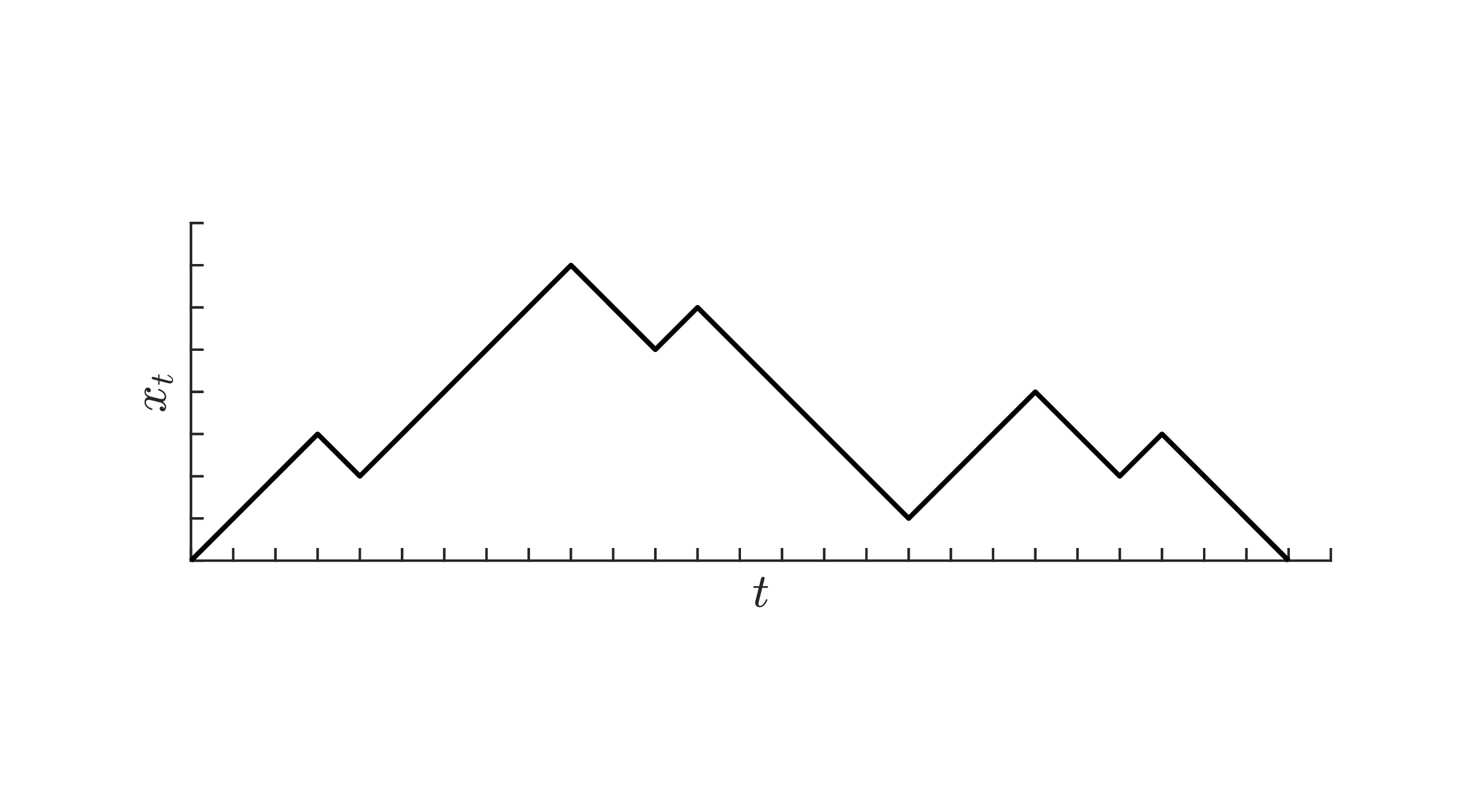}
}
\caption{
Illustration of a first return trajectory of an RW on a one-dimensional lattice. 
It shows the location $x_t$ 
of the RW as a function of time,
starting from the origin at $t=0$ until it returns to the origin for the first time. 
The number $s$ of distinct sites visited by the RW before returning
to the origin (not including the origin itself) is given by the maximum value of $x_t$.
}
\label{fig:1}
\end{figure}

In Fig. \ref{fig:1} we present an illustration of a first return trajectory of an RW
on a one-dimensional lattice. It shows the location $x_t$ of the RW as a
function of time, starting from the origin $x_0=0$ at time $t=0$, until it returns to the origin for the first time
at time $t$. The number $s$ of distinct sites visited by the RW before returning
to the origin (not including the origin itself) is given by the maximum height of the trajectory,
or the maximum distance
from the origin that the RW has reached along its path.
Such trajectory is referred to as a random walk bridge
\cite{Godreche2015}.

In order to derive an equation for the joint distribution 
$P(T_{\rm FR}=t,S=s)$
of first return times and of the number of distinct sites visited by
an RW before returning to the origin, one needs more
detailed combinatorial properties of the first return trajectories.
More specifically, one needs to evaluate the combinatorial factor
$T(n,s)$, which accounts for the number of RW trajectories 
on the positive half of the lattice that 
reach a maximum distance $s$ before returning
to the origin at time $t=2n$.
The combinatorial factor $T(n,s)$ includes trajectories that return to the origin at earlier times 
$t'=2n'$, where $0 < n' < n$,
but does not include trajectories that cross the origin to the negative side.
Since the RW needs to reach a distance $s$ and return to the origin at time $t=2n$,
it is clear that $T(n,s)=0$ for $s>n$.
It is also clear that there is a single path for which $s=n$, namely $T(n,n)=1$ for $n \ge 0$.
The number of RW trajectories that reach a maximum distance of $s$ before 
returning to the origin for the {\it first} time at $t=2n$ is given by 
$T(n-1,s-1)$. This is due to the fact that in these trajectories the
first step is always to the right and the last step is always to the left, thus the
remaining RW trajectory is of length
$t=2(n-1)$ and the maximum distance is $s-1$.

The joint distribution of first return times and of the number of distinct sites visited by
an RW before returning to the origin is given by

\begin{equation}
P(T_{\rm FR}=2n,S=s) =
T(n-1,s-1) \left( \frac{1}{2} \right)^{2n-1},
\label{eq:PTFRnh}
\end{equation}

\noindent
where
$T(n,s)$ accounts for the number of RW trajectories of length $t=2n$ that 
reach a maximum distance of $s$ and may visit the origin but do not cross it to the negative side, before returning
to the origin at time $t=2n$.
Such trajectories are known as bounded Dyck paths
\cite{Flajolet2009,Hein2022}.

Note that the initial site $x_0=0$ is not counted, namely $s$ counts the number
of distinct sites visited at times $1 \le t' \le 2n-1$.
The combinatorial factor $T(n,s)$ can be expressed in the form

\begin{equation}
T(n,s) = U(n,s) - U(n,s-1),
\label{eq:Tnh}
\end{equation}

\noindent
where $U(n,s)$ is the number of RW trajectories of length $2n$ that
reach a maximum distance which is smaller or equal to $s$,
before returning to the origin (not necessarily for the first time)
and do not cross to the negative side.

The combinatorial factor $U(n,s)$ was derived in Refs.
\cite{Bruijn1972,Hein2022}
and is given by

\begin{equation}
U(n,s) = \frac{2^{2n+1}}{s+2} \sum_{m=1}^{s+1} 
\sin^2 \left( \frac{ m \pi }{s+2}   \right) 
\cos^{2n} \left( \frac{ m \pi }{s+2} \right).
\label{eq:Snh}
\end{equation}

\noindent
Since the largest possible number of distinct sites visited by an RW trajectory that returns to the 
origin at time $t=2n$ is $s=n$, we conclude that $U(n,n)$ accounts for all the RW trajectories
that do not cross the origin to the negative side and
return to the origin at time $t=2n$. This implies that $U(n,n)=C_n$, where $C_n$ is
the Catalan number,
given by Eq. (\ref{eq:Catalan}). 
This equality is shown explicitly in Appendix A.

Inserting $T(n,s)$ from Eq. (\ref{eq:Tnh}) into Eq. (\ref{eq:PTFRnh}) 
and using $U(n,s)$ from Eq. (\ref{eq:Snh}),
we obtain

\begin{eqnarray}
P(T_{\rm FR}=2n,S=s) &=&
\frac{1}{s+1} \sum_{m=1}^{s} 
\sin^2 \left( \frac{ m \pi }{s+1}   \right) 
\cos^{2n-2} \left( \frac{ m \pi }{s+1} \right) 
\nonumber \\
&-&
\frac{1}{s} \sum_{m=1}^{s-1} 
\sin^2 \left( \frac{ m \pi }{s}   \right) 
\cos^{2n-2} \left( \frac{ m \pi }{s} \right).
\label{eq:PTFRnh2b}
\end{eqnarray}

The generating functions of $U(n,s)$
with respect to $n$ is given by

\begin{equation}
E_s(y) = \sum_{n=0}^{\infty} y^n U(n,s).
\label{eq:Ehy}
\end{equation}

\noindent
Inserting $U(n,s)$ from Eq. (\ref{eq:Snh}) into Eq. (\ref{eq:Ehy})
and carrying out the summation, we obtain

\begin{equation}
E_s(y) = 2 \frac{ \left( 1 + \sqrt{1-4y} \right)^{s+1} - \left( 1 - \sqrt{1-4y} \right)^{s+1} }
{ \left( 1 + \sqrt{1-4y} \right)^{s+2} - \left( 1 - \sqrt{1-4y} \right)^{s+2} }.
\label{eq:Ehy2}
\end{equation}

\noindent
This generating function corresponds to the probability generating function presented in equation (6)
of Klinger et al. 
\cite{Klinger2022}.

Below we show that the combinatorial factor $U(n,s)$ satisfies the recursion equation

\begin{equation}
U(n+1,s) = \sum_{i=0}^n U(i,s-1) U(n-i,s).
\label{eq:Un1sRE}
\end{equation}

\noindent
Multiplying Eq. (\ref{eq:Un1sRE}) by $y^{n+1}$ and rearranging terms, one obtains

\begin{equation}
E_{s}(y) - U(0,s) = y E_{s-1}(y) E_{s}(y).
\label{eq:Esx}
\end{equation}

\noindent
Inserting $U(0,s)=1$ in Eq. (\ref{eq:Esx}) and rearranging terms, we obtain
a recursion equation for $E_s(y)$, which takes the form

\begin{equation}
E_s(y) = \frac{1}{1-y E_{s-1}(y)}.
\label{eq:EsxRE}
\end{equation}

\noindent
Inserting $E_s(y)$ from Eq. (\ref{eq:Ehy2}) into Eq. (\ref{eq:EsxRE}),
one can show that $U(n,s)$ indeed satisfies the
recursion equation (\ref{eq:Un1sRE}).

Using the trigonometric identity 
$\sin^2 (\alpha) = 1 - \cos^2 (\alpha)$, 
Eq. (\ref{eq:Snh}) can be 
expressed in the form

\begin{equation}
U(n,s) =  2^{2n+1} 
\left[ 
F(n,s) - F(n+1,s)
\right],
\label{eq:Snh2}
\end{equation}

\noindent
where

\begin{equation}
F(n,s) = \frac{1}{s+2} 
\sum_{m=0}^{s+1} 
\cos^{2n} \left( \frac{ m \pi }{s+2} \right).
\label{eq:Fnh}
\end{equation}

\noindent
In Appendix A we summarize some useful properties of $F(n,s)$ and provide alternative expressions
in terms of binomial coefficients, based on Refs. 
\cite{Prudnikov1998,Merca2012,Merca2013,Merca2014,Fonseca2017}.
Expressing the right hand side of Eq. (\ref{eq:PTFRnh2b}) in terms of the function $F(n,s)$,
we obtain

\begin{equation}
P(T_{\rm FR}=2n,S=s) = 
F(n-1,s-1) - F(n,s-1) - F(n-1,s-2) + F(n,s-2).
\end{equation}

The distribution of the number of distinct sites visited by an RW before returning to the
origin for the first time is given by

\begin{equation}
P(S=s) = \sum_{n=s}^{\infty} P(T_{\rm FR}=2n,S=s).
\label{eq:PHh}
\end{equation}
 
\noindent
Inserting $P(T_{\rm FR}=2n,S=s)$ from Eq. (\ref{eq:PTFRnh2b}) into Eq. (\ref{eq:PHh}),
one obtains

\begin{eqnarray}
P(S=s) &=& \frac{1}{s+1} \sum_{m=1}^{s} 
\sin^2 \left( \frac{m \pi}{s+1} \right)
\sum_{n=s}^{\infty} \cos^{2n-2} \left( \frac{m \pi}{s+1} \right)
\nonumber \\
&-&
\frac{1}{s} \sum_{m=1}^{s-1} 
\sin^2 \left( \frac{m \pi}{s} \right)
\sum_{n=s}^{\infty} \cos^{2n-2} \left( \frac{m \pi}{s} \right).
\label{eq:PHh2}
\end{eqnarray}

\noindent
The summations over $n$ in Eq. (\ref{eq:PHh2}) are simply geometric series,
which can be easily calculated.
Carrying out the summations over $n$,
one obtains

\begin{equation}
P(S=s) = 
F(s-1,s-1) - F(s-1,s-2),
\label{eq:PHh4}
\end{equation}

\noindent
where $F(n,s)$ is given by Eq. (\ref{eq:Fnh}).
Using Eq. (\ref{eq:Fnsbc2}) in Appendix A to evaluate both terms on the right hand side of Eq. (\ref{eq:PHh4}),
we recover the known result 
\cite{Lawler2010}

\begin{equation}
P(S=s) = \frac{1}{s(s+1)}.
\label{eq:PHh7}
\end{equation}

\noindent
This is a fat-tailed power-law distribution in which RW trajectories that
visit a very large number of distinct sites carry much weight.
This can be attributed to the fact that
the distribution of first return times is fat-tailed and to
the strong correlation between the first return time of an RW trajectory and the
number of distinct sites visited by the RW.
In fact, the mean number of distinct sites 
$\langle S \rangle$ visited by an RW before returning to the
origin diverges logarithmically.

From Eq. (\ref{eq:PHh7}), one can obtain the
tail distribution of the number of distinct sites visited by an RW
before returning to the origin, which is given by
$P(S \ge s) = 1/s$.
This result is in agreement with Proposition 5.1.1 in Ref.
\cite{Lawler2010}, regarding the Gambler's ruin estimate.
To illustrate this point consider a gambler who starts with an initial fortune
of one dollar, and on each successive round either wins one dollar or loses
one dollar with equal probabilities, until he/she gets ruined (runs out of cash). 
In this context, $P(S \ge s)$ is the probability that the gambler will amass at least
$s$ dollars at some point in the game before getting ruined.

In finite systems the number of distinct sites visited before an RW hits a target
is bounded by the system size. In the case of a finite one-dimensional lattice the
distribution $P(S=s)$ was calculated in Ref.
\cite{Klinger2021}.

\section{The conditional distribution $P(T_{\rm FR}=2n | S=s)$}

The conditional distribution of first return times of RWs under the condition that the number of
distinct sites visited before returning to the origin is $s$, is given by

\begin{equation}
P(T_{\rm FR}=2n | S=s) =
\frac{ P(T_{\rm FR}=2n, S=s) }{ P(S=s) }.
\label{eq:PTFRH}
\end{equation}

\noindent
Inserting 
$P(T_{\rm FR}=2n, S=s)$
from Eq. (\ref{eq:PTFRnh})
and 
$P(S=s)$
from Eq. (\ref{eq:PHh7})
into Eq. (\ref{eq:PTFRH}),
we obtain

\begin{equation}
P(T_{\rm FR}=2n | S=s) =
s(s+1) T(n-1,s-1) \left( \frac{1}{2} \right)^{2n-1}.
\label{eq:PTFRH2}
\end{equation}

\noindent
Inserting $T(n,s)$ from Eq. (\ref{eq:Tnh}) into Eq. (\ref{eq:PTFRH2}),
we obtain

\begin{equation}
P(T_{\rm FR}=2n | S=s) =
s(s+1) \left[ U(n-1,s-1) - U(n-1,s-2) \right] \left( \frac{1}{2} \right)^{2n-1}.
\label{eq:PTFRH3}
\end{equation}

\noindent
Inserting $U(n,s)$ from Eq. (\ref{eq:Snh}) into Eq. (\ref{eq:PTFRH3}), we obtain

\begin{eqnarray}
P(T_{\rm FR}=2n | S=s) &=&
s
\sum_{m=1}^{s} 
\sin^2 \left( \frac{ m \pi }{s+1}   \right) 
\cos^{2n-2} \left( \frac{ m \pi }{s+1} \right).    
\nonumber \\
&-&
(s+1)     
\sum_{m=1}^{s-1} 
\sin^2 \left( \frac{ m \pi }{s}   \right) 
\cos^{2n-2} \left( \frac{ m \pi }{s} \right).   
\label{eq:PTFRH4}
\end{eqnarray}

\begin{figure}
\centerline{
\includegraphics[width=7cm]{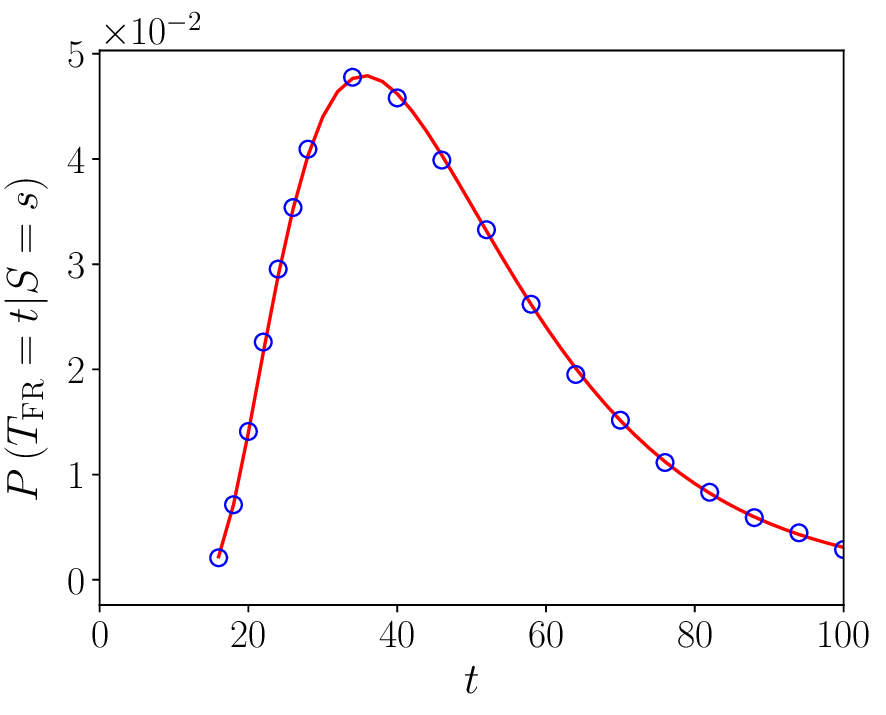}
}
\caption{
Analytical results (solid line) for the conditional probability
distribution  
$P(T_{\rm FR}=t | S=s)$  
of first return times $t=2n$ of RWs that have visited $s=8$ distinct sites
before returning to the origin for the first time.
The analytical results,
obtained from Eq. (\ref{eq:PTFRH4}),
are in very good agreement with the results obtained
from computer simulations (circles).
}
\label{fig:2}
\end{figure}

In Fig. \ref{fig:2} we present analytical
results (solid line) for the conditional probability
distribution  
$P(T_{\rm FR}=t | S=s)$  
of first return times of RWs that have visited $s=8$ distinct sites 
before returning to the origin for the first time.
The analytical results,
obtained from Eq. (\ref{eq:PTFRH4}),
are in very good agreement with the results obtained
from computer simulations (circles).
The simulation results presented in this paper are based on a set of $10^7$
first return trajectories.
The number of first return trajectories for which $s=8$ is $N(s=8) = 138,544$.
Using standard methods for the analysis of statistical errors, it is found that the error bars
for the simulation results are negligibly small.

In the asymptotic long time limit, where $n \gg s$, the tail of 
$P(T_{\rm FR}=2n | S=s)$
decays exponentially according to

\begin{equation}
P(T_{\rm FR}=2n | S=s)  \simeq
2 s
\tan^2 \left( \frac{  \pi }{s+1}   \right) 
\cos^{2n} \left( \frac{  \pi }{s+1} \right).    
\label{eq:PTFRH4as}
\end{equation}

\section{The mean and variance of $P(T_{\rm FR}=2n | S=s)$}

Below we calculate the moments of the conditional distribution
$P(T_{\rm FR}=2n | S=s)$, denoted by

\begin{equation}
{\mathbb E}[T_{\rm FR}^{r} | S=s] =
\sum_{n=1}^{\infty} (2n)^r P(T_{\rm FR}=2n | S=s).
\end{equation}

\noindent
To this end, we use the moment generating function

\begin{equation}
R_s(x) =\sum_{n=1}^{\infty}  e^{2 n x} P(T_{\rm FR}=2n | S=s).
\label{eq:Rh}
\end{equation}

\noindent
Inserting $P(T_{\rm FR}=2n | S=s)$ from Eq. (\ref{eq:PTFRH2})
into Eq. (\ref{eq:Rh}), we obtain

\begin{equation}
R_s(x) = s(s+1) \frac{e^{2x}}{2} 
\sum_{n=0}^{\infty} T(n,s-1) \left( \frac{e^x}{2} \right)^{2n}.
\label{eq:Rh2}
\end{equation}

\noindent
Using Eq. (\ref{eq:Tnh}), we rewrite Eq. (\ref{eq:Rh2}) in the form

\begin{equation}
R_s(x) = s(s+1) \frac{ e^{2x} }{2} 
\left[ 
\sum_{n=0}^{\infty} U(n,s-1) \left( \frac{ e^x }{2} \right)^{2n}
-
\sum_{n=0}^{\infty} U(n,s-2) \left( \frac{ e^x }{2} \right)^{2n}
\right].
\label{eq:Rhx}
\end{equation}

\noindent
The right hand side of 
Eq. (\ref{eq:Rhx}) can be expressed in terms of the generating function
$E_s(y)$, defined in Eq. (\ref{eq:Ehy}).
Eq. (\ref{eq:Rhx}) can thus be rewritten in the form

\begin{equation}
R_s(x) = s(s+1) \frac{ e^{2x} }{2} 
\left[ 
E_{s-1} \left( \frac{ e^{2x} }{4} \right)
-
E_{s-2} \left( \frac{ e^{2x} }{4} \right)
\right].
\label{eq:Rhx2}
\end{equation}

\noindent
The $r$th moment of $P(T_{\rm FR}=2n | S=s)$ can be expressed
in terms of the $r$th derivative of $R_s(x)$, in the form

\begin{equation}
{\mathbb E}[T_{\rm FR}^{r} | S=s] =
\frac{d^r R_s(x)}{dx^r} \bigg\vert_{x=0}.
\label{eq:ETFRrh}
\end{equation}

\noindent
Inserting $R_s(x)$ from Eqs. (\ref{eq:Rhx2}) and (\ref{eq:Ehy2}) into
the right hand side of Eq. (\ref{eq:ETFRrh}) and evaluating the first derivative ($r=1$)
at $x=0$, we obtain
the first moment of the conditional distribution, which is given by

\begin{equation}
{\mathbb E}[T_{\rm FR}| S=s] =
\frac{2}{3} \left( s^2 + s + 1 \right).
\label{eq:ETFRrh1}
\end{equation}

\noindent
Evaluating the second derivative ($r=2$) on the right hand side of Eq. (\ref{eq:ETFRrh}),
we obtain

\begin{equation}
{\mathbb E}[T_{\rm FR}^2| S=s] =
\frac{4}{45} \left( 6 s^4 + 12 s^3 + 13 s^2 + 7 s + 7  \right).
\label{eq:ETFRrh2}
\end{equation}

\begin{figure}
\centerline{
\includegraphics[width=7cm]{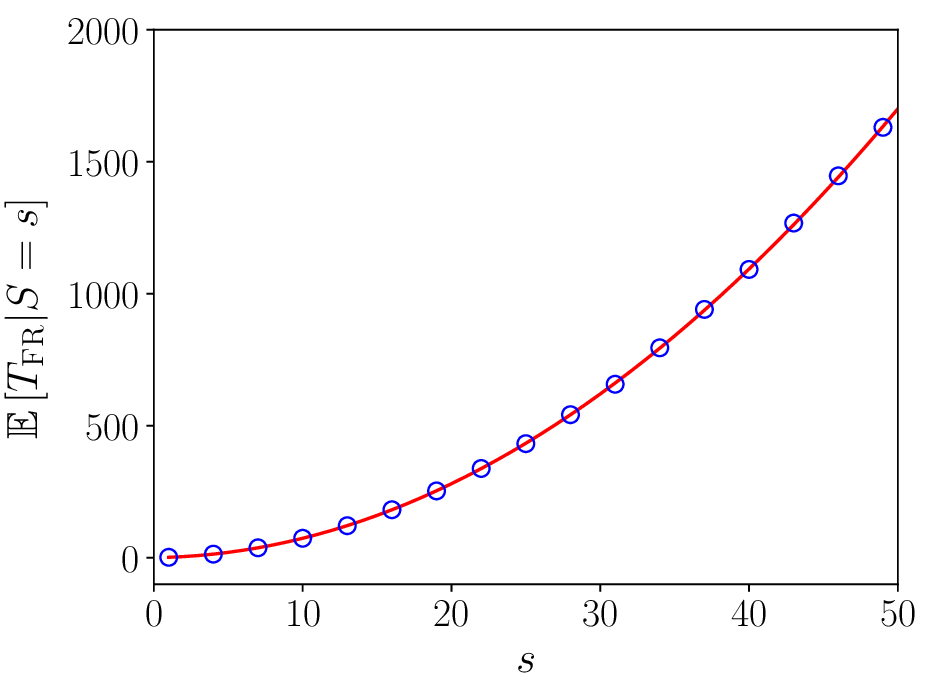}
}
\caption{
Analytical results (solid line) for the  
expectation value 
${\mathbb E}[T_{\rm FR} | S=s]$
of the first return time 
of RWs that have visited $s$ distinct sites 
before returning to the origin for the first time,
as a function of $s$.
The analytical results, 
obtained from Eq. (\ref{eq:ETFRrh1}),
are in very good agreement with the results obtained
from computer simulations (circles).
}
\label{fig:3}
\end{figure}

In Fig. \ref{fig:3} we present analytical
results (solid line) for the  
expectation value 
${\mathbb E}[T_{\rm FR} | S=s]$
of the first return time 
of RWs that have visited $s$ distinct sites 
before returning to the origin for the first time.
The analytical results,
obtained from Eq. (\ref{eq:ETFRrh1}),
are in very good agreement with the results obtained
from computer simulations (circles).

The variance of the conditional probability distribution is given by

\begin{equation}
{\rm Var}(T_{\rm FR} | S=s) =
{\mathbb E}[T_{\rm FR}^2| S=s]
-
\left( {\mathbb E}[T_{\rm FR}| S=s] \right)^2.
\label{eq:VarTFRHh}
\end{equation}

\noindent
Inserting
${\mathbb E}[T_{\rm FR}| S=s]$
from Eq. (\ref{eq:ETFRrh1}) 
and
${\mathbb E}[T_{\rm FR}^2| S=s]$
from Eq. (\ref{eq:ETFRrh2})
into Eq. (\ref{eq:VarTFRHh}),
we obtain

\begin{equation}
{\rm Var}(T_{\rm FR} | S=s) =
\frac{4}{45} (s-1)(s+2)(s^2+s-1).
\label{eq:VarTFRHh2}
\end{equation}

\begin{figure}
\centerline{
\includegraphics[width=7cm]{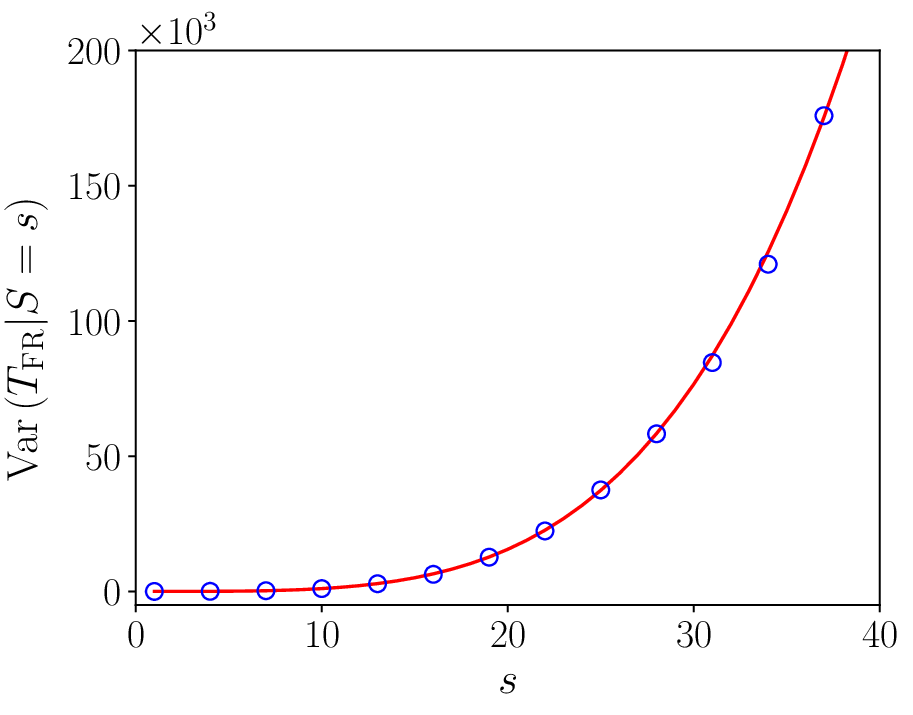}
}
\caption{
Analytical results (solid line) for the  
variance 
${\rm Var}(T_{\rm FR} | S=s)$
of the conditional distribution of first return times 
of RWs that have visited $s$ distinct sites 
before returning to the origin for the first time,
as a function of $s$.
The analytical results, 
obtained from Eq. (\ref{eq:VarTFRHh2}),
are in very good agreement with the results obtained
from computer simulations (circles).
}
\label{fig:4}
\end{figure}

In Fig. \ref{fig:4} we present
analytical results (solid line) for the  
variance 
${\rm Var}(T_{\rm FR} | S=s)$
of the conditional distribution of first return times
of RWs that have visited $s$ distinct sites 
before returning to the origin for the first time,
as a function of $s$.
The analytical results,
obtained from Eq. (\ref{eq:VarTFRHh2}),
are in very good agreement with the results obtained
from computer simulations (circles).

\section{The conditional distribution $P(S=s | T_{\rm FR}=2n)$}

The conditional probability that an RW visited $s$ distinct sites,
before returning to the origin for the first time at $t=2n$, 
is given by

\begin{equation}
P(S=s | T_{\rm FR}=2n) =
\frac{ P(T_{\rm FR}=2n, S=s) }{ P(T_{\rm FR}=2n) }.
\label{eq:PHhTFRn}
\end{equation}

\noindent
Inserting 
$P(T_{\rm FR}=t, S=s)$
from Eq. (\ref{eq:PTFRnh})
and 
$P(T_{\rm FR}=2n)$
from Eq. (\ref{eq:PTFRR})
into Eq. (\ref{eq:PHhTFRn}),
we obtain

\begin{equation}
P(S=s | T_{\rm FR}=2n) =
\frac{ T(n-1,s-1) }{ C_{n-1} }.
\label{eq:PHhTFRn2}
\end{equation}

\noindent
Inserting $T(n-1,s-1)$ from Eq. (\ref{eq:Tnh}) into Eq. (\ref{eq:PHhTFRn2}),
we obtain

\begin{equation}
P(S=s | T_{\rm FR}=2n) =
\frac{ 1 }{ C_{n-1} }
\left[ U(n-1,s-1) - U(n-1,s-2) \right].
\label{eq:PHhTFRn3}
\end{equation}

\noindent
Inserting $U(n,s)$ from Eq. (\ref{eq:Snh2}) into Eq. (\ref{eq:PHhTFRn3}), we obtain

\begin{eqnarray}
P(S=s | T_{\rm FR}=2n) &=&
\frac{ 2^{2n+1} }{ C_{n-1} }
\left[ F(n,s) - F(n+1,s) 
\right.
\nonumber \\
&-& \left. F(n,s-1) + F(n+1,s-1) \right],
\label{eq:PHhTFRn3b}
\end{eqnarray}

\noindent
where $F(n,s)$ may be expressed either by
Eq. (\ref{eq:Fnh}), 
Eq. (\ref{eq:Fnsbc1}) 
or by Eq. (\ref{eq:Fnsbc2}).

\begin{figure}
\centerline{
\includegraphics[width=7cm]{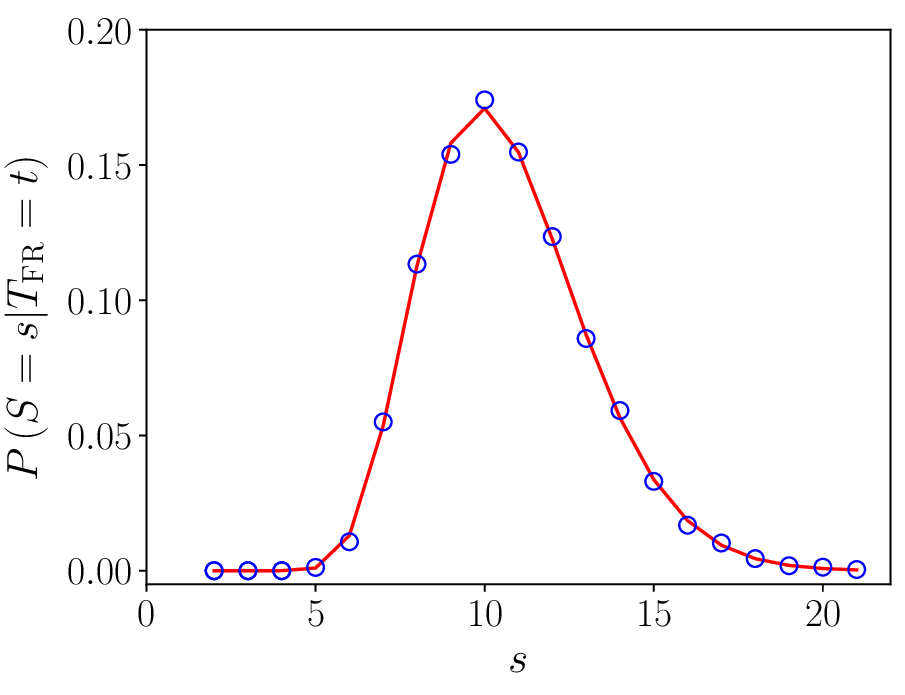}
}
\caption{
Analytical results (solid line) for the conditional probability
distribution  
$P(S=s | T_{\rm FR}=t)$  
of the number of distinct sites visited by RWs
whose first return time to the origin is $t=80$.
The analytical results,
obtained from Eq. (\ref{eq:PHhTFRn2}),
are in very good agreement with the results obtained
from computer simulations (circles).
}
\label{fig:5}
\end{figure}

In Fig. \ref{fig:5} we present analytical
results (solid line) for the conditional probability 
distribution  
$P(S=s | T_{\rm FR}=t)$  
of the number of distinct sites visited by RWs
whose first return time to the origin is $t=80$.
The analytical results,
obtained from Eq. (\ref{eq:PHhTFRn2}),
are in very good agreement with the results obtained
from computer simulations (circles).
In the simulation results, the number of first return trajectories for which
$t=80$ is $N(t=80) = 11,413$.
Using standard methods for the analysis of statistical errors, it is found that the error bars
for the simulation results are negligibly small.

The cumulative distribution $P(S \le s | T_{\rm FR}=2n)$ is given by

\begin{equation}
P(S \le s | T_{\rm FR}=2n) = \sum_{s'=0}^{s} P(S = s' | T_{\rm FR}=2n).
\label{eq:PSTcum} 
\end{equation}

\noindent
Inserting $P(S = s | T_{\rm FR}=2n)$ from Eq. (\ref{eq:PHhTFRn2}) into Eq. (\ref{eq:PSTcum}),
we obtain

\begin{equation}
P(S \le s | T_{\rm FR}=2n) =
\frac{ U(n-1,s-1) }{ C_{n-1} }.
\label{eq:PHhTFRn5}
\end{equation}

An asymptotic analysis of $P(S \le s | T_{\rm FR}=2n)$,
in the large $n$ limit is presented on page 329 in 
Ref. \cite{Flajolet2009}.
In this analysis $s$ is expressed in the form
$s=x \sqrt{n}$,
where $x$ is a continuous variable.
It is shown that in the large $n$ limit the random variable $S/\sqrt{n}$ 
obeys a Theta distribution, expressed by

\begin{equation}
P \left(\frac{S}{\sqrt{n}} \le x \right) = \frac{ 4 \pi^{5/2} }{x^3}
\sum_{j=0}^{\infty} j^2 e^{-\pi^2 j^2 / x^2}.
\label{eq:thetadist}
\end{equation}

\section{The mean and variance of $P(S=s | T_{\rm FR}=2n)$}

Below we calculate the first two moments of the conditional distribution
$P(S=s | T_{\rm FR}=2n)$, denoted by

\begin{equation}
{\mathbb E}[S^r | T_{\rm FR}=2n ] =
\sum_{s=1}^{n} s^r P(S=s | T_{\rm FR}=2n),
\label{eq:EHrTFR2n}
\end{equation}

\noindent
where $r=1$ and $2$, respectively.
Inserting $P(S=s | T_{\rm FR}=2n)$ from Eq. (\ref{eq:PHhTFRn2})
into Eq. (\ref{eq:EHrTFR2n}), we obtain

\begin{equation}
{\mathbb E}[S^r | T_{\rm FR}=2n ] =
\frac{1}{C_{n-1}} \sum_{s=1}^{n} s^r T(n-1,s-1).
\label{eq:EHrTFR2n2}
\end{equation}

\noindent
Inserting $T(n-1,s-1)$ from Eq. (\ref{eq:Tnh}) into Eq. (\ref{eq:EHrTFR2n2}),
we obtain

\begin{equation}
{\mathbb E}[S^r | T_{\rm FR}=2n ] =
\frac{1}{C_{n-1}} \sum_{s=1}^{n} s^r  U(n-1,s-1)
-
\frac{1}{C_{n-1}} \sum_{s=2}^{n} s^r  U(n-1,s-2).
\label{eq:EHrTFR2n3}
\end{equation}

\noindent
Inserting $U(n,s)$ from Eq. (\ref{eq:Snh}) into Eq. (\ref{eq:EHrTFR2n3}),
we obtain

\begin{eqnarray}
&{\mathbb E}[S^r | T_{\rm FR}=2n ]  = 
\nonumber \\
&
\ \ \ \ \ \ \ \ 
\frac{2^{2n-1}}{C_{n-1}} \sum_{s=1}^{n} \frac{s^r}{s+1}
\sum_{m=1}^{s} \left[ \cos^{2n-2} \left( \frac{m \pi}{s+1} \right)
- \cos^{2n} \left( \frac{m \pi}{s+1} \right) \right]
\nonumber \\
&
\ \ \ \ \ \ \   
-
\frac{2^{2n-1}}{C_{n-1}} \sum_{s=2}^{n} \frac{s^r}{s}
\sum_{m=1}^{s-1} \left[ \cos^{2n-2} \left( \frac{m \pi}{s} \right)
- \cos^{2n} \left( \frac{m \pi}{s} \right) \right]. 
\label{eq:EHrTFR2n4}
\end{eqnarray}

\noindent
Eq. (\ref{eq:EHrTFR2n4}) provides the conditional moment 
${\mathbb E}[S^r|T_{\rm FR}=2n]$
in terms of two double sums.
Each double sum includes $n^2/2$ terms,
where each term includes differences of high powers of
trigonometric functions.
This combination makes the evaluation of 
the sums very difficult when $n$ becomes large.
One difficulty is due to the mere complexity, namely the number
of steps in the calculation. 
Another difficulty stems from the extremely high precision 
that is required for the evaluation of the differences between
the terms in the square brackets.
In practice, the evaluation of 
${\mathbb  E}[S^r|T_{\rm FR}=2n]$
via Eq. (\ref{eq:EHrTFR2n4})
for $n=1000$ is already challenging.

In Appendix B, we derive
expressions for the first two conditional moments
${\mathbb  E}[S^r|T_{\rm FR}=2n]$, $r=1,2$,
which are much easier to evaluate
than Eq. (\ref{eq:EHrTFR2n4}).
This is done by expressing the right hand side of Eq. (\ref{eq:EHrTFR2n4})
in terms of binomial coefficients.
For the first moment ($r=1$),
it yields

\begin{eqnarray}
&{\mathbb E}[S | T_{\rm FR}=2n ]    =   
7 - \frac{30 n}{(n+1)(n+2)}
- \frac{n}{\binom{2n-2}{n-1}}
\times
\nonumber \\
&
\ \ \ \ \ \ \ \ 
\sum_{s=1}^{ \left\lfloor \frac{n-3}{3} \right\rfloor }
\left[  
4   \sum_{k=3}^{\left\lfloor \frac{n-1}{s+1} \right\rfloor}
\binom{2n-2}{n-1+k(s+1)}  
-
\sum_{k=3}^{\left\lfloor \frac{n}{s+1} \right\rfloor}
\binom{2n}{n+k(s+1)}  
\right],
\label{eq:EHrTFR2n15}
\end{eqnarray}

\noindent
where $\lfloor x \rfloor$ is the integer part of $x$.
Eq. (\ref{eq:EHrTFR2n15}) provides the conditional mean
${\mathbb E}[S | T_{\rm FR}=2n ]$
in terms of binomial coefficients, which are easier to
evaluate than the high powers of trigonometric functions,
which appear on the right hand side of Eq. (\ref{eq:EHrTFR2n9}).
Using Eq. (\ref{eq:EHrTFR2n15}) we managed to perform direct numerical
evaluation of 
${\mathbb  E}[S|T_{\rm FR}=2n]$
with $n=10^6$.

\begin{figure}
\centerline{
\includegraphics[width=7cm]{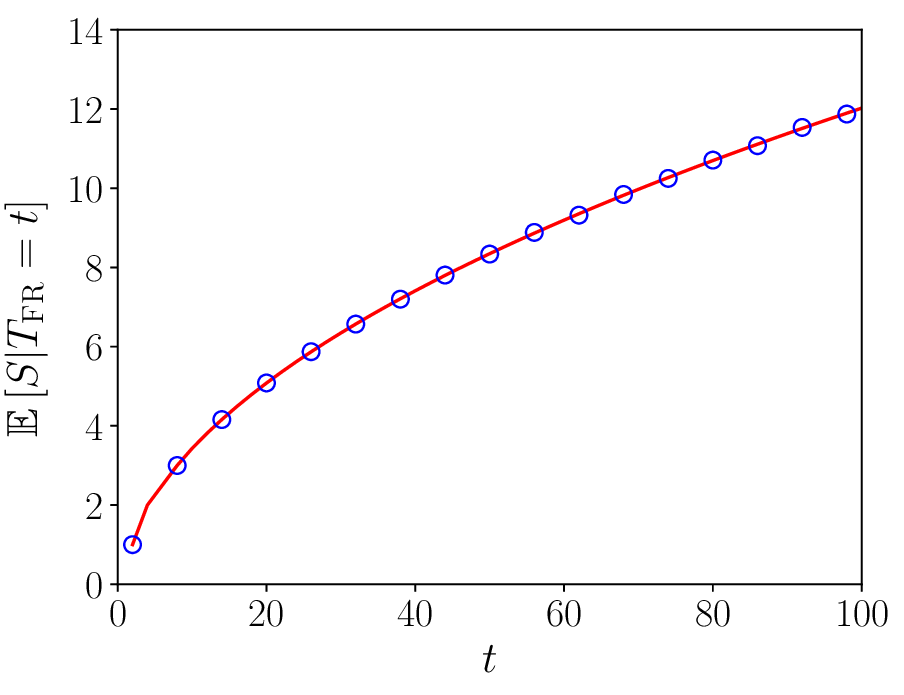}
}
\caption{
Analytical results (solid line) for the conditional
expectation value 
${\mathbb E}[S|T_{\rm FR}=t]$
of the number of distinct sites visited by an RW 
whose first return time to the origin is $t$.
The analytical results,
obtained from Eq. (\ref{eq:EHrTFR2n15}),
are in very good agreement with the results obtained
from computer simulations (circles).
}
\label{fig:6}
\end{figure}

In Fig. \ref{fig:6} we present analytical
results (solid line) for the conditional
expectation value 
${\mathbb E}[S|T_{\rm FR}=t]$
of the number of distinct sites visited by RWs
whose first return time to the origin is $t$.
The analytical results,
obtained from Eq. (\ref{eq:EHrTFR2n15}),
are in very good agreement with the results obtained
from computer simulations (circles).

Using direct numerical evaluation of the right hand side of  
Eq. (\ref{eq:EHrTFR2n15}), with very high precision,
and fitting the results to the anticipated scaling form 
in the limit of large $n$,
it is found that the tail of
${\mathbb E}[S | T_{\rm FR}=2n ]$ satisfies

\begin{equation}
{\mathbb E}[S | T_{\rm FR}=2n ] \simeq
\sqrt{ \frac{\pi}{2} } \sqrt{2n}.
\label{eq:EStail}
\end{equation}

\noindent
The great precision that can be achieved using Eq. (\ref{eq:EHrTFR2n15})
is the key to being able to obtain this asymptotic result.

In order to explore the convergence of
${\mathbb E}[S | T_{\rm FR}=2n ]$
towards its asymptotic form in the large $n$ limit,
we examine the difference

\begin{equation}
\Delta_1 = 1 -  \frac{ {\mathbb E}[S | T_{\rm FR}=2n ] }{ \sqrt{ \frac{\pi}{2} } \sqrt{2n} },
\label{eq:Delta1}
\end{equation}

\noindent
where  ${\mathbb E}[S | T_{\rm FR}=2n ]$
is evaluated by Eq. (\ref{eq:EHrTFR2n15}).

\begin{figure}
\centerline{
\includegraphics[width=7cm]{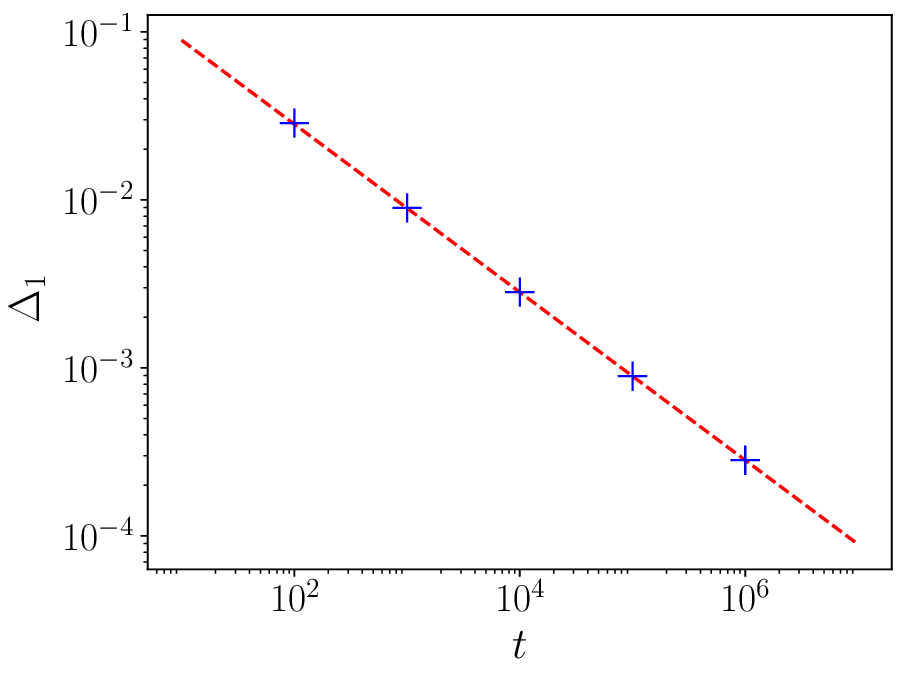}
}
\caption{
Analytical results ($+$ symbols) for the difference
$\Delta_1$, given by Eq. (\ref{eq:Delta1}),
as a function of the time $t$.
These results are obtained by direct numerical evaluation of
${\mathbb E}[S | T_{\rm FR}=t ]$, using Eq. (\ref{eq:EHrTFR2n15}).
These results are well fitted by Eq. (\ref{eq:Delta1b}), 
represented by the dashed line.
We thus conclude that the tail of 
${\mathbb E}[S | T_{\rm FR}=t ]$
follows Eq. (\ref{eq:EStail}), namely
${\mathbb E}[S | T_{\rm FR}=t ] \simeq \sqrt{ \frac{\pi}{2} } \sqrt{t}$.
}
\label{fig:7}
\end{figure}

In Fig. \ref{fig:7} we present
analytical results ($+$ symbols) for the difference $\Delta_1$,
as a function of the time $t=2n$.
The analytical results are obtained from Eq. (\ref{eq:Delta1}),
where the conditional expectation value
${\mathbb E}[S | T_{\rm FR}=t ]$ 
is calculated by direct numerical evaluation of
Eq. (\ref{eq:EHrTFR2n15}).
These results are well fitted by 

\begin{equation}
\Delta_1 = \frac{1}{2 \sqrt{\pi n} },
\label{eq:Delta1b}
\end{equation}

\noindent
which is shown by a dashed line.
This confirms that the tail of ${\mathbb E}[S | T_{\rm FR}=t ]$
satisfies Eq. (\ref{eq:EStail}), where $t=2n$, and suggests the next order, namely

\begin{equation}
{\mathbb E}[S | T_{\rm FR}=t ] \simeq \sqrt{ \frac{\pi}{2} } \sqrt{t} - \frac{1}{2}.
\end{equation}

\noindent
This result is consistent with Eq. (69) in Chapter V (page 329) of Ref. \cite{Flajolet2009}.

To evaluate the second conditional moment
${\mathbb E}[S^2 | T_{\rm FR}=2n ]$,
given by Eq. (\ref{eq:EHrTFR2n4}),
in Appendix B we express it in terms of
binomial coefficients. 
This yields

\begin{eqnarray}
&{\mathbb E}[S^2 | T_{\rm FR}=2n ]   =   
\frac{5}{2} n + \frac{41}{4} + \frac{60}{n+1} - \frac{120}{n+2} + \frac{3}{4(2n-3)}
\nonumber \\
&
- {\mathbb E}[S | T_{\rm FR}=2n ]
-
\frac{2n}{\binom{2n-2}{n-1}} \sum_{s=1}^{\left\lfloor \frac{n-3}{3} \right\rfloor} 
(s+1)
\times
\nonumber \\
&
\ \ \ \ \ \ \ \ \ \ 
\left[
4 \sum_{k=3}^{ \left\lfloor \frac{n-1}{s+1} \right\rfloor }
\binom{2n-2}{n-1+k(s+1)}
-
\sum_{k=3}^{ \left\lfloor \frac{n}{s+1} \right\rfloor }
\binom{2n}{n+k(s+1)}  
\right].   
\label{eq:EH5}
\end{eqnarray}

\noindent
Using direct numerical evaluation of the right hand side of  
Eq.  (\ref{eq:EH5}), with very high precision,
and fitting the results to the anticipated scaling form 
in the limit of large $n$,
it is found that the tail of
${\mathbb E}[S^2 | T_{\rm FR}=2n ]$ 
satisfies

\begin{equation}
{\mathbb E}[S^2 | T_{\rm FR}=2n ] \simeq
\frac{\pi^2}{3} n - \sqrt{\pi n}.
\label{eq:ES2tail}
\end{equation}

The leading order of all the moments of $P(S=s|T_{\rm FR}=2n)$ is presented in
Eq. (70) in Chapter V (page 329) of Ref. \cite{Flajolet2009}, and is given by

\begin{equation}
{\mathbb E}[S^r | T_{\rm FR}=2n ] \simeq
[r(r-1) \Gamma(r/2) \zeta(r)] n^{r/2},
\label{eq:SrTFR}
\end{equation}

\noindent
where $\Gamma(x)$ is the gamma function and $\zeta(x)$ is the Riemann zeta function
\cite{Olver2010}.
The leading order on the right hand side of Eq. (\ref{eq:ES2tail}) is in agreement with
Eq. (\ref{eq:SrTFR}), where $r=2$.
However, the subleading term on the right hand side of Eq. (\ref{eq:ES2tail}) is new.

In order to explore the convergence of
${\mathbb E}[S^2 | T_{\rm FR}=2n ]$
towards its asymptotic form in the large $n$ limit,
we examine the difference

\begin{equation}
\Delta_2 = 1 -  \frac{ {\mathbb E}[S^2 | T_{\rm FR}=2n ] }{ \frac{\pi^2}{3} n - \sqrt{\pi n}  },
\label{eq:Delta2}
\end{equation}

\noindent
where  ${\mathbb E}[S^2 | T_{\rm FR}=2n ]$
is evaluated by Eq. (\ref{eq:EH5}).

\begin{figure}
\centerline{
\includegraphics[width=7cm]{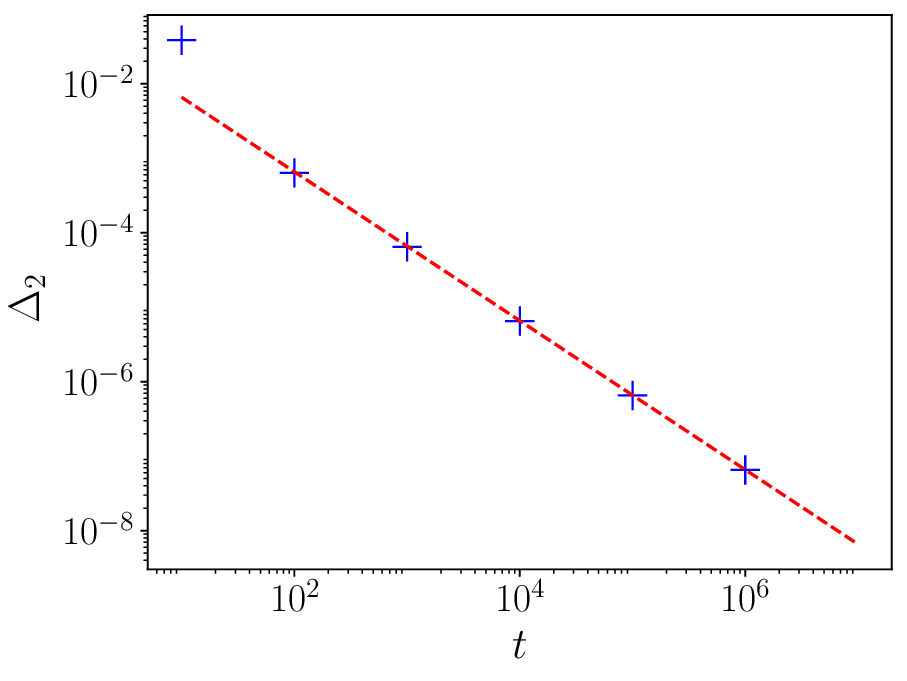}
}
\caption{
Analytical results ($+$ symbols) for the difference
$\Delta_2$, given by Eq. (\ref{eq:Delta2}),
as a function of the time $t$.
The analytical results are
obtained by direct numerical evaluation of
${\mathbb E}[S^2 | T_{\rm FR}=t ]$, using Eq. (\ref{eq:EH5}).
These results are well fitted by Eq. (\ref{eq:Delta2b}),
represented by the dashed line.
We thus conclude
that the tail of ${\mathbb E}[S^2 | T_{\rm FR}=t ]$
follows Eq. (\ref{eq:ES2tail}), namely
${\mathbb E}[S | T_{\rm FR}=t ] \simeq \frac{\pi^2}{6} t - \sqrt{\frac{\pi}{2}} \sqrt{t}$.
}
\label{fig:8}
\end{figure}

In Fig. \ref{fig:8} we present
analytical results ($+$ symbols) for the difference
$\Delta_2$, given by Eq. (\ref{eq:Delta2}),
as a function of $t=2n$.
The analytical results are obtained by direct numerical evaluation of
${\mathbb E}[S^2 | T_{\rm FR}=t ]$, using Eq. (\ref{eq:EH5}).
These results are well fitted by 

\begin{equation}
\Delta_2 = \frac{\pi^2}{150 \, n}, 
\label{eq:Delta2b}
\end{equation}

\noindent
which is shown by a dashed line.
This confirms that the tail of ${\mathbb E}[S^2 | T_{\rm FR}=t ]$
follows Eq. (\ref{eq:ES2tail}), where $t=2n$.
It also suggests the form of the next subleading term, implying that

\begin{equation}
{\mathbb E}[S^2 | T_{\rm FR}=t ] \simeq \frac{\pi^2}{6} t 
- \sqrt{\frac{\pi}{2}} \sqrt{t} - \frac{\pi^4}{450}.
\label{eq:ES3tail}
\end{equation}

\noindent
Note that unlike the leading term
on the right hand side of Eq. (\ref{eq:ES3tail}),
which appeared in Ref. \cite{Flajolet2009},
the two subleading terms were not known before.

The variance of $P(S=s | T_{\rm FR}=t)$ is given by

\begin{equation}
{\rm Var}(S|T_{\rm FR}=t) =
{\mathbb E}[S^2|T_{\rm FR}=t]
-
\left({\mathbb E}[S|T_{\rm FR}=t]\right)^2.
\label{eq:VarST}
\end{equation}

\begin{figure}
\centerline{
\includegraphics[width=7cm]{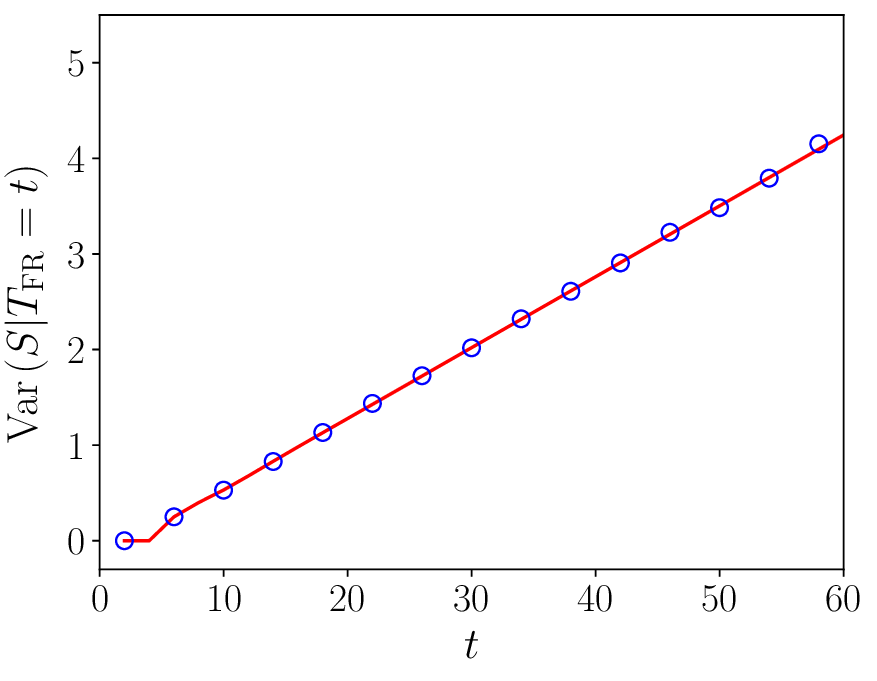}
}
\caption{
Analytical results (solid line) for the  
variance
${\rm Var}(S|T_{\rm FR}=t)$
of the distribution of the number of distinct sites visited by an RW 
whose first return time to the origin is $t$.
The analytical results, 
obtained from Eq. (\ref{eq:VarST}),
are in very good agreement with the results obtained
from computer simulations (circles).
}
\label{fig:9}
\end{figure}

\noindent
In Fig. \ref{fig:9} we present 
analytical results (solid line) for the  
variance
${\rm Var}(S|T_{\rm FR}=t)$
of the distribution of the number of distinct sites visited by an RW 
whose first return time to the origin is $t$.
The analytical results are
obtained from Eq. (\ref{eq:VarST}),
where 
${\mathbb E}[S^2|T_{\rm FR}=t]$
is given by Eq. (\ref{eq:EH5})
and
${\mathbb E}[S|T_{\rm FR}=t]$
is given by Eq. (\ref{eq:EHrTFR2n15}).
They are in very good agreement with the results obtained
from computer simulations (circles).

\section{Discussion}

In this paper we address the relation between the first return time of an RW trajectory and
the number of distinct sites visited by the RW before returning to the origin. 
To this end, we consider the joint distribution $P(T_{\rm FR}=t,S=s)$, from which we
extract the conditional distributions $P(T_{\rm FR}=t|S=s)$ and $P(S=s|T_{\rm FR}=t)$.
The importance of the joint distribution $P(T_{\rm FR}=t,S=s)$ 
stems from the fact that it
captures the interplay between the kinetic and
the geometric properties of the first return trajectories
\cite{Klinger2022}.
Unlike the marginal distributions $P(T_{\rm FR}=t)$ and $P(S=s)$, which exhibit diverging first moments,  
the two conditional distributions that emerge from $P(T_{\rm FR}=t,S=s)$
exhibit finite moments. 
As a result, they provide a solid theoretical framework, which is crucial for the study of 
the kinetic and geometric properties of first return trajectories.

Using a generating function approach, we calculate the conditional expectation value
${\mathbb E}[T_{\rm FR}|S=s]$
and the variance
${\rm Var}(T_{\rm FR}|S=s)$.
In the large $s$ limit it is found that
${\mathbb E}[T_{\rm FR}|S=s] \simeq \frac{2}{3} s^2$
and
${\rm Var}(T_{\rm FR}|S=s) \simeq \frac{4}{45} s^4$.
We also calculate the conditional expectation value
${\mathbb E}[S|T_{\rm FR}=t]$
and the variance
${\rm Var}(S|T_{\rm FR}=t)$.
In the limit of large $t$ it is found that
${\mathbb E}[S|T_{\rm FR}=t] \simeq  \sqrt{\frac{\pi}{2}} \sqrt{t} + \mathcal{O}(1)$
and the variance
${\rm Var}(S|T_{\rm FR}=t) \simeq \left( \frac{\pi^2}{6} - \frac{\pi}{2} \right) t
-\sqrt{\frac{\pi}{2}} \sqrt{t} + \mathcal{O}(1)$.

It is interesting to compare the conditional expectation value
${\mathbb E}[S|T_{\rm FR}=t]$
of the number of distinct sites visited by an RW before returning to the origin for the first
time at time $t$,
given by Eq. (\ref{eq:EStail}), 
and the 
mean number of distinct sites $\langle S \rangle_t$ visited by an RW up to time $t$,
given by Eq. (\ref{eq:Smean1D}).
Both quantities scale like $\sqrt{t}$, however the pre-factor of
${\mathbb E}[S|T_{\rm FR}=t]$
is smaller than the pre-factor of
$\langle S \rangle_t$.
This appears to be due to the condition that first return trajectories are required
to return to the origin at time $t$. As a result, for a period of time before returning
to the origin, the RW must retrace its path backwards toward the origin and during
this time it cannot visit new sites.
This is unlike the unconditioned RW trajectories that may visit new sites at any time
step up to time $t$.

The symmetric RW studied in this paper is a special case of a more general RW model
in which at each time step the RW moves to the right with probability $p$ or to the 
left with probability $q=1-p$. In case that $p>1/2$ the RW is biased to the right,
while in case that $p<1/2$ the RW is biased to the left. Here we focus on the case in
which in the first step the RW moves to the right, namely $x_1=1$.
In case that $p>1/2$ the RW becomes transient and the return probability to
the origin is $P_{\rm R}= (1-p)/p$.
In case that $p<1/2$ the RW is recurrent, namely $P_{\rm R}=1$ and the
mean first return time becomes finite and is given by
$\langle T_{\rm FR} \rangle = 2(1-p)/(1-2p)$.
The case of $p=q=1/2$ is special in the sense that in this case the RW is recurrent
but the mean first return time to the origin diverges.
Results for the joint generating function of biased RWs were recently derived in Ref.
\cite{Klinger2022}.

In recent years there has been much interest in resetting RWs
\cite{Evans2011}.
At each time step a resetting RW
may either hop randomly to one of the neighboring sites or 
all the way to the origin (a resetting step). 
As a result, in a resetting RW the first return process may occur either
via normal diffusion or via resetting.
Resetting RWs on a one dimensional lattice are recurrent, and unlike the case ordinary RWs
the mean first return time of a resetting RW is finite.
Results for the joint generating function for RWs with resetting were recently
presented in Ref. \cite{Klinger2022}.

\section{Summary}

We presented analytical results for the joint distribution 
$P(T_{\rm FR}=t,S=s)$
of FR times $t$ and of the number of distinct sites $s$  
visited by an RW on a one dimensional lattice
before returning to the origin.
These results provide a formulation that controls the divergences
in the distributions $P(T_{\rm FR}=t)$ and $P(S=s)$
and accounts for the interplay between 
the kinetic and geometric properties of first return trajectories.
We calculated the conditional distributions 
$P(T_{\rm FR}= t|S=s)$
and
$P (S=s|T_{\rm FR}=t)$.
Using moment generating functions and combinatorial methods, 
we found that the conditional expectation value 
of first return times of trajectories that visit $s$ distinct sites is
${\mathbb E}[T_{\rm FR} | S=s] = \frac{2}{3} (s^2+s+1)$,
and the variance is
${\rm Var}(T_{\rm FR} | S=s)=\frac{4}{45} (s-1)(s+2)(s^2+s-1)$.
We also found that in the asymptotic limit, the conditional expectation value of the
number of distinct sites visited by an RW that first returns to the origin at time $t=2n$ is
${\mathbb E}[S|T_{\rm FR}=2n] \simeq \sqrt{ \pi n }$,
and the variance is
${\rm Var}(S|T_{\rm FR}=2n) \simeq  \pi \left( \frac{\pi}{3} - 1 \right)  n$.
The analytical results were
found to be in very good agreement with the results 
obtained from computer simulations.
These results go beyond the important recent results of Klinger et al. 
\cite{Klinger2022},
who derived a closed form expression for the generating function of the
joint distribution, but did not go further to extract an explicit expression for
the joint distribution itself.

It will be interesting to generalize the analysis to a broader class of random walk models,
which includes biased RWs and resetting RWs.
In biased RWs on a one-dimensional lattice, the probability $p$ to hop to the right
and the probability $q=1-p$ to hop to the left are not the same, making the 
calculations more difficult.
In resetting RWs one must distinguish between two types of first return processes,
namely an ordinary diffusive return to the origin and return due to resetting.

A further challenge will be to extend this analysis to higher-dimensional lattices
and to complex networks, where the first return trajectories exhibit complex geometries.
Unlike the one dimensional case, in which the number of distinct sites $s$ coincides with
the maximum distance $h$ from the origin, in higher dimensions these two quantities
are not the same.
In the case of complex networks, the joint distribution will provide useful insight on the efficiency of   
search, sampling and exploration processes using RWs
\cite{Cooper2016,Katzir2011}, 
in which the aim is to cover 
as many distinct sites as possible in a given number of steps.

\appendix

\section{Properties of $F(n,s)$}

In this Appendix we analyze the properties of the combinatorial coefficient
$F(n,s)$, given by Eq. (\ref{eq:Fnh}).
In the special case of $s=n$, one can show that
\cite{Prudnikov1998}

\begin{equation}
F(n,n) = \binom{2n}{n} \frac{1}{2^{2n}}.
\label{eq:Fnn}
\end{equation}

\noindent
Inserting $F(n,n)$ from Eq. (\ref{eq:Fnn}) into Eq. (\ref{eq:Snh2}) it is found that
indeed $U(n,n) = C_n$, namely it captures all the first return trajectories of length $t=2n$.
More surprisingly, it turns out that for any values of $s \ge n$,
$F(n,s)$ is also given by
\cite{Prudnikov1998}

\begin{equation}
F(n,s) = \binom{2n}{n} \frac{1}{2^{2n}},
\label{eq:Fnns2}
\end{equation}

\noindent
namely it does not depend on $s$.
The case of $s > n$ is not relevant to first return trajectories because the largest
distance that can be reached by an RW in such trajectory is $s=n$.
In fact, in case that $s > n$ the sum expressed by $F(n,s)$ is equal to the corresponding 
integral, namely

\begin{equation}
F(n,s) = \int_{0}^{1} \cos^{2n} (\pi x) dx.
\end{equation}

\noindent
This implies that when $s > n$ the Riemann sum, which consist of intervals whose width is $1/(s+2)$,
is sufficiently refined that it is coincides with the corresponding integral. 
In other words, it implies that in case that $s>n$ the corrections associated with the Euler-Maclaurin
expansion vanish.

More generally, it can be shown that for any combination of $n$ and $s$,
the combinatorial factor $F(n,s)$ can be expressed in terms of binomial
coefficients in the form
\cite{Merca2012,Merca2013,Merca2014,Fonseca2017}

\begin{equation}
F(n,s) = \frac{1}{2^{2n}}
\sum_{k= - \left\lfloor \frac{n}{s+2} \right\rfloor}^{\left\lfloor \frac{n}{s+2} \right\rfloor}
\binom{2n}{n+k(s+2)}.
\label{eq:Fnsbc1}
\end{equation}

\noindent
In order to distinguish between the case of $s+2 \le n$ and the
case of $s+2 > n$ in a more transparent way, Eq. (\ref{eq:Fnsbc1}) can be written in the form

\begin{equation}
F(n,s) = 
\left\{
\begin{array}{ll}
\frac{1}{2^{2n}} \binom{2n}{n} & n<s+2 \\
\\
\frac{1}{2^{2n}} \binom{2n}{n} +
\frac{1}{2^{2n-1}}
\sum\limits_{k=1}^{\left\lfloor \frac{n}{s+2} \right\rfloor}
\binom{2n}{n+k(s+2)}
& n \ge s+2
\end{array}
\right.
\label{eq:Fnsbc2}
\end{equation}

\noindent
Inserting the right hand side of Eq. (\ref{eq:Fnsbc2}) into Eq. (\ref{eq:Snh2}),
the combinatorial factor $U(n,s)$ can be expressed in the form

\begin{eqnarray}
U(n,s) =
\frac{1}{2} \sum_{k= - \left\lfloor \frac{n}{s+2} \right\rfloor }^{\left\lfloor \frac{n}{s+2} \right\rfloor} 
\left[ 4 \binom{2n}{n+k(s+2)} - \binom{2n+2}{n+1+k(s+2)} \right].
\label{eq:Unsbinom}
\end{eqnarray}

\noindent
For large values of $n$, the numerical evaluation of $U(n,s)$ may become 
computationally heavy. In case that $U(n,s)$ is evaluated via Eq. 
(\ref{eq:Snh}), the number of terms in the summation is $s$,
while in case that it is evaluated via Eq. (\ref{eq:Unsbinom}) the
number of terms is of order $\lfloor n/s \rfloor$.
This implies that for $s < \sqrt{n}$ it is more efficient to evaluate
$U(n,s)$ via Eq. (\ref{eq:Snh}), while for $s > \sqrt{n}$ it is more
efficient to use Eq. (\ref{eq:Unsbinom}).

\section{Efficient evaluation of the moments ${\mathbb E}[S^r|T_{\rm FR}=2n]$}

In this Appendix we derive expressions for 
the conditional moments
${\mathbb E}[S^r | T_{\rm FR}=2n ]$, where $r=1$ and $2$,
which are easier to calculate by direct numerical evaluation
than Eq. (\ref{eq:EHrTFR2n15}).
To this end, we replace the double sums over high powers of trigonometric functions
by double sums of binomial coefficients.
The latter sums consist of a smaller number of terms, 
which can be evaluated more efficiently and to higher precision.

Shifting the summation index in the second term of Eq. (\ref{eq:EHrTFR2n4})
from $s$ to $s-1$, we obtain

\begin{eqnarray}
&{\mathbb E}[S^r | T_{\rm FR}=2n ]  = 
\label{eq:EHrTFR2n5}
\\
&
\ \ \ \ \ \ \ \ 
\frac{2^{2n-1}}{C_{n-1}} \sum_{s=1}^{n} \frac{s^r}{s+1}
\sum_{m=1}^{s} \left[ \cos^{2n-2} \left( \frac{m \pi}{s+1} \right)
- \cos^{2n} \left( \frac{m \pi}{s+1} \right) \right]
\nonumber \\
&
\ \ \ \ \ \ \ 
-
\frac{2^{2n-1}}{C_{n-1}} \sum_{s=1}^{n-1} \frac{(s+1)^r}{s+1}
\sum_{m=1}^{s} \left[ \cos^{2n-2} \left( \frac{m \pi}{s+1} \right)
- \cos^{2n} \left( \frac{m \pi}{s+1} \right) \right]. 
\nonumber
\end{eqnarray}

\noindent
Setting the upper limits of both summations at $s=n-1$ and presenting separately the
$s=n$ term from the first summation, we obtain

\begin{eqnarray}
&{\mathbb E}[S^r | T_{\rm FR}=2n ]  = 
\nonumber \\
&
\ \ \ \ 
\frac{2^{2n-1}}{C_{n-1}} \sum_{s=1}^{n-1} 
\frac{s^r - (s+1)^r}{s+1}
\sum_{m=1}^{s} \left[ \cos^{2n-2} \left( \frac{m \pi}{s+1} \right)
- \cos^{2n} \left( \frac{m \pi}{s+1} \right) \right]
\nonumber \\
&
\ \ \ 
+
\frac{2^{2n-1}}{C_{n-1}}   \frac{n^r}{n+1}
\sum_{m=1}^{n} \left[ \cos^{2n-2} \left( \frac{m \pi}{n+1} \right)
- \cos^{2n} \left( \frac{m \pi}{n+1} \right) \right]. 
\label{eq:EHrTFR2n6}
\end{eqnarray}

\noindent
Expressing the second term 
on the right hand side of Eq. (\ref{eq:EHrTFR2n6}) in terms of
$F(n,s)$, given by Eq. (\ref{eq:Fnh}), we obtain

\begin{eqnarray}
&{\mathbb E}[S^r | T_{\rm FR}=2n ]  = 
\nonumber \\
&
\ \ \ \ 
\frac{2^{2n-1}}{C_{n-1}} \sum_{s=1}^{n-1} 
\frac{s^r - (s+1)^r}{s+1}
\sum_{m=1}^{s} \left[ \cos^{2n-2} \left( \frac{m \pi}{s+1} \right)
- \cos^{2n} \left( \frac{m \pi}{s+1} \right) \right]
\nonumber \\
&
\ \ \ 
+
\frac{2^{2n-1}}{C_{n-1}}    n^r 
\left[ F(n-1,n-1) - F(n,n-1) \right].
\label{eq:EHrTFR2n7}
\end{eqnarray}

\noindent
Evaluating the second term on the right hand side of Eq. (\ref{eq:EHrTFR2n7}),
using Eq. (\ref{eq:Fnsbc2}), 
we obtain

\begin{eqnarray}
{\mathbb E}[S^r | T_{\rm FR}=2n ]  = & 
n^r +
\frac{2^{2n-1}}{C_{n-1}} \sum_{s=1}^{n-1} 
\frac{s^r - (s+1)^r}{s+1}
\times
\nonumber \\
&
\ \ \ \ \ \ \ \ 
\sum_{m=1}^{s}  \sin^{2} \left( \frac{m \pi}{s+1} \right)  \cos^{2n-2} \left( \frac{m \pi}{s+1} \right).
\label{eq:EHrTFR2n8}
\end{eqnarray}

\noindent
Inserting $r=1$ in Eq. (\ref{eq:EHrTFR2n8}), we obtain an expression for the expected number
of distinct nodes visited by an RW in a first return trajectory of length $t=2n$.
It is given by

\begin{equation}
{\mathbb E}[S | T_{\rm FR}=2n ]  = 
n -
\frac{2^{2n-1}}{C_{n-1}} 
\sum_{s=1}^{n-1} 
\frac{1}{s+1}
\sum_{m=1}^{s}  \sin^{2} \left( \frac{m \pi}{s+1} \right)  \cos^{2n-2} \left( \frac{m \pi}{s+1} \right).
\label{eq:EHrTFR2n9}
\end{equation}

\noindent
Using a trigonometric identity, we write Eq. (\ref{eq:EHrTFR2n9}) in the form

\begin{equation}
{\mathbb E}[S | T_{\rm FR}=2n ]  = 
n -
\frac{2^{2n-1}}{C_{n-1}} 
\sum_{s=1}^{n-1} 
\left[ F(n-1,s-1) - F(n,s-1) \right].
\label{eq:EHrTFR2n10}
\end{equation}

\noindent
Inserting $F(n,s)$ from Eq. (\ref{eq:Fnsbc2}) into Eq. (\ref{eq:EHrTFR2n10}),
we obtain

\begin{eqnarray}
&{\mathbb E}[S | T_{\rm FR}=2n ]   =  
\nonumber \\
&
n -
\frac{n}{2  \binom{2n-2}{n-1}} 
\sum_{s=1}^{n-1} 
\left[  
4 \binom{2n-2}{n-1} - \binom{2n}{n} +
8 \sum_{k=1}^{ \left\lfloor \frac{n-1}{s+1} \right\rfloor }
\binom{2n-2}{n-1+k(s+1)}
\right.
\nonumber \\
&
-
\left.
2 \sum_{k=1}^{\left\lfloor \frac{n}{s+1} \right\rfloor}
\binom{2n}{n+k(s+1)}
\right].
\label{eq:EHrTFR2n12}
\end{eqnarray}

\noindent
Rearranging terms, we obtain

\begin{eqnarray}
&{\mathbb E}[S | T_{\rm FR}=2n ]   =  1 -
\frac{n}{ \binom{2n-2}{n-1} } \times
\nonumber \\
&
\sum_{s=1}^{n-1} 
\left[  
4   \sum_{k=1}^{\left\lfloor \frac{n-1}{s+1} \right\rfloor}
\binom{2n-2}{n-1+k(s+1)}  
\right.
- 
\left.
\sum_{k=1}^{\left\lfloor \frac{n}{s+1} \right\rfloor}
\binom{2n}{n+k(s+1)}  
\right].
\label{eq:EHrTFR2n13}
\end{eqnarray}

\noindent
Separating the $s=n-1$ term and the $k=1$ term from the rest of the sum,
we obtain

\begin{eqnarray}
&{\mathbb E}[S | T_{\rm FR}=2n ]   =  
1 
+ \frac{n}{\binom{2n-2}{n-1}}
\nonumber \\
& 
\ 
- \frac{n}{ \binom{2n-2}{n-1} }
\sum_{s=1}^{n-2}
\left[   4 \binom{2n-2}{n-1+(s+1)}  
-
   \binom{2n}{n+(s+1)}  
\right]
\label{eq:EHrTFR2n14}
\\
&- 
\frac{n}{ \binom{2n-2}{n-1} } 
\sum_{s=1}^{ \left\lfloor \frac{n-2}{2} \right\rfloor } 
\left[  
4   \sum_{k=2}^{\left\lfloor \frac{n-1}{s+1} \right\rfloor}
\binom{2n-2}{n-1+k(s+1)}  
\right.
-
\left.
\sum_{k=2}^{\left\lfloor \frac{n}{s+1} \right\rfloor}
\binom{2n}{n+k(s+1)}  
\right].    
\nonumber
\end{eqnarray}

\noindent
Evaluating the first three terms, we obtain

\begin{eqnarray}
&{\mathbb E}[S | T_{\rm FR}=2n ]   =  
4 - \frac{6}{n+1}
\label{eq:EHrTFR2n14b}
\\
&- 
\frac{n}{ \binom{2n-2}{n-1} } 
\sum_{s=1}^{ \left\lfloor \frac{n-2}{2} \right\rfloor } 
\left[  
4   \sum_{k=2}^{\left\lfloor \frac{n-1}{s+1} \right\rfloor}
\binom{2n-2}{n-1+k(s+1)}  
\right.
-
\left.
\sum_{k=2}^{\left\lfloor \frac{n}{s+1} \right\rfloor}
\binom{2n}{n+k(s+1)}  
\right].
\nonumber
\end{eqnarray}

\noindent
Separating the $s=(n-2)/2$ term and the $k=2$ term from the rest of the sum,
we obtain

\begin{eqnarray}
&{\mathbb E}[S | T_{\rm FR}=2n ]    =   
7 - \frac{30 n}{(n+1)(n+2)}
\label{eq:EHrTFR2n15B}
\\
&
- \frac{n}{\binom{2n-2}{n-1}}
\sum_{s=1}^{ \left\lfloor \frac{n-3}{3} \right\rfloor }
\left[  
4   \sum_{k=3}^{\left\lfloor \frac{n-1}{s+1} \right\rfloor}
\binom{2n-2}{n-1+k(s+1)}  
-
\sum_{k=3}^{\left\lfloor \frac{n}{s+1} \right\rfloor}
\binom{2n}{n+k(s+1)}  
\right].
\nonumber
\end{eqnarray}

\noindent
Eq. (\ref{eq:EHrTFR2n15B}) provides a much more efficient evaluation of
${\mathbb E}[S | T_{\rm FR}=2n ]$
than Eq. (\ref{eq:EHrTFR2n4}) with $r=1$.

Below we present a derivation, whose aim is to obtain an expression for the second moment
${\mathbb E}[S^2 | T_{\rm FR}=2n ]$,
which is amenable to efficient numerical evaluation.
Inserting
$r=2$ in Eq. (\ref{eq:EHrTFR2n8}),
we obtain

\begin{equation}
{\mathbb E}[S^2 | T_{\rm FR}=2n ]  = 
n^2 -
\frac{2^{2n-1}}{C_{n-1}} \sum_{s=1}^{n-1} 
\frac{2s+1}{s+1}
\sum_{m=1}^{s}  \sin^{2} \left( \frac{m \pi}{s+1} \right)  \cos^{2n-2} \left( \frac{m \pi}{s+1} \right).
\label{eq:EH2}
\end{equation}
\noindent

\noindent
Splitting the sum on the right hand side of Eq. (\ref{eq:EH2}) into two separate terms, 
using the fact that $2s+1 = 2(s+1)-1$,
we obtain

\begin{eqnarray}
&{\mathbb E}[S^2 | T_{\rm FR}=2n ]   =  
n^2 -
2 \frac{2^{2n-1}}{C_{n-1}} \sum_{s=1}^{n-1} 
\sum_{m=1}^{s}  \sin^{2} \left( \frac{m \pi}{s+1} \right)  \cos^{2n-2} \left( \frac{m \pi}{s+1} \right)
\nonumber \\
&
\ \ \ \ \ \ \ \ \ \ \ \ 
+ 
\frac{2^{2n-1}}{C_{n-1}} \sum_{s=1}^{n-1} 
\frac{1}{s+1}
\sum_{m=1}^{s}  \sin^{2} \left( \frac{m \pi}{s+1} \right)  \cos^{2n-2} \left( \frac{m \pi}{s+1} \right).
\label{eq:EH3}
\end{eqnarray}

\noindent
Using Eq. (\ref{eq:EHrTFR2n9}), we express the third term on the right hand side of Eq. (\ref{eq:EH3})
in terms of ${\mathbb E}[S | T_{\rm FR}=2n ]$, and obtain

\begin{eqnarray}
{\mathbb E}[S^2 | T_{\rm FR}=2n ]  &=&
n^2 -
2 \frac{2^{2n-1}}{C_{n-1}} \sum_{s=1}^{n-1} 
\sum_{m=1}^{s}  \sin^{2} \left( \frac{m \pi}{s+1} \right)  \cos^{2n-2} \left( \frac{m \pi}{s+1} \right)
\nonumber \\
&+&
\left\{
n - {\mathbb E}[S | T_{\rm FR}=2n ]
\right\}.
\label{eq:EH4}
\end{eqnarray}

\noindent
Expressing the sum over trigonometric functions
on the right hand side of Eq. (\ref{eq:EH4}) in terms of sums over 
binomial coefficients and rearranging terms, 
we obtain

\begin{eqnarray}
&{\mathbb E}[S^2 | T_{\rm FR}=2n ]   =   
\frac{5}{2} n + \frac{41}{4} + \frac{60}{n+1} - \frac{120}{n+2} + \frac{3}{4(2n-3)}
\nonumber \\
&- {\mathbb E}[S | T_{\rm FR}=2n ]
-
\frac{2n}{\binom{2n-2}{n-1}} \sum_{s=1}^{\left\lfloor \frac{n-3}{3} \right\rfloor} 
(s+1)
\times
\nonumber \\
&
\ \ \ \ \ \ \ \ 
\left[
4 \sum_{k=3}^{ \left\lfloor \frac{n-1}{s+1} \right\rfloor }
\binom{2n-2}{n-1+k(s+1)}
-
\sum_{k=3}^{ \left\lfloor \frac{n}{s+1} \right\rfloor }
\binom{2n}{n+k(s+1)}  
\right].   
\label{eq:EH5B}
\end{eqnarray}

\noindent
Similar steps could be used to derive simplified expressions for higher moments as well.

\noappendix

\end{document}